\documentstyle[sprocl]{article}

\def\Journal#1#2#3#4{{#1} {\bf #2}, #3 (#4)}

\def\NPB{{\it Nucl. Phys.} B}
\def\PLB{{\it Phys. Lett.}  B}

\def\be{\begin{equation}}
\def\ee{\end{equation}}

\def\d3r{\delta^{(3)}({\bf r})} 
\def\vol{{\sf V}}
\def\num{{\sf N}}
\def\half{{\scriptstyle \frac{1}{2}}}

\begin{document}

\title{ARE THERE PRESSURE WAVES IN THE VACUUM?}

\author{P. M. STEVENSON}

\address{T. W. Bonner Laboratory, Department of Physics and Astronomy,\\ 
Rice University, PO Box 1892, Houston, \\ TX 77251, USA\\
E-mail: stevenson@physics.rice.edu}

\maketitle\abstracts{The Higgs vacuum is a kind of medium.  In any 
medium one generally expects sound waves for sufficiently long 
wavelengths ($\lambda \gg$ mean free path).  I briefly describe how the 
broken-symmetry vacuum can be viewed as a Bose-Einstein condensate of 
`phion' particles.  This picture yields a natural notion of the `mean 
free path.'  I speculate that this is at the millimeter-centimeter scale. }

\section{Introduction}

     {\it Are there sound waves in the vacuum?} The question might seem  
silly, but here's my point:  (i) the Higgs vacuum of the Standard Model, 
with its non-zero background field $\langle \Phi \rangle \neq 0$, is not 
`empty,' but rather is a kind of medium.  (ii) For very basic reasons, one 
expects any medium to propagate pressure waves for sufficiently long 
wavelengths.  At length scales much, much longer than the mean free path 
({\it mfp}) --- the `hydrodynamical' regime --- sound waves arise directly 
from energy-momentum conservation equations linearized about 
equilibrium.\cite{lif,huang}  Could this reasoning apply even to the 
`aether' (the Higgs vacuum) --- and, if so, what is the `{\it mfp}' scale?  

     In this talk I briefly outline a description~\cite{mech} of the Higgs 
vacuum as, literally, a Bose-Einstein (BE) condensate of particles that
I'll call `{\it phions}.'  I'll consider single-component $\lambda \Phi^4$ 
theory since Goldstone bosons are not directly relevant here.  The intrinsic 
phion size $r_0$ will serve as the inverse of the ultraviolet cutoff.  The 
key variables will be $n$, the number density of phions, and $a$, their 
scattering length.  

     In the quantum-field-theory limit, where $r_0 \to 0$, it will turn out 
that $n \to \infty$ and $a \to 0$, such that $na$ remains finite and 
determines the Higgs mass squared.  The fact that the scattering length $a$ 
goes to zero reflects the `triviality' of $\lambda \Phi^4$ theory.~\cite{triv} 
In this field-theory limit sound waves would not exist; they would be 
banished to infinitely long wavelengths because the {\it mfp} scale is of 
order $1/(na^2)$ which goes to infinity.   

     However, if we do not take the cutoff all the way to infinity then the 
{\it mfp} $1/(na^2)$ would be a large, but finite, length scale.  If we are 
also willing to contemplate Lorentz-invariance violation at the cutoff scale, 
we could naturally expect sound waves to exist at ultra-low momenta.  They 
would be an example of what Volovik~\cite{volovik} (in a different context) 
has called ``re-entrant violation of Lorentz invariance,'' in which Lorentz 
invariance arises as a low-energy effective symmetry from a non-symmetric 
fundamental theory, but deviations from Lorentz symmetry occur at 
ultra-{\it low} momenta as well as at very high momenta.  I'll return to these 
speculations later.

\section{Phion condensation}

      Consider single-component $\lambda \Phi^4$ theory in a region of 
parameters where the effective potential has both a minimum at $\phi=0$ 
and a deeper minimum at $\phi=\pm v$.  (Such a situation {\it is} possible, 
as we'll see.)  The symmetric vacuum is then locally, but not globally, 
stable.  The particle excitations above this metastable, `empty' vacuum I 
will call `phions.'  It must be possible to describe the physics in terms of 
the phion degrees of freedom.  The broken-symmetry vacuum must correspond
to a BE condensate of phions, and the Higgs bosons must correspond to the 
{\it quasiparticle} excitations of this condensate, in condensed-matter 
terminology.  The issue, though, is; why do phions want to condense?

      The answer lies, of course, in the phions' interactions.  The 
fundamental interaction is the 4-point vertex.  Expressed as an interparticle 
potential this is a repulsive $\d3r$ potential with strength $a/m$,\footnote{
Throughout this talk I'll ignore numerical factors of 2, $4\pi$, etc. }
where $m$ is the phion mass and $a$ is the scattering length 
($a\sim\lambda/m$ in terms of the coupling constant $\lambda$).  The $\d3r$ 
potential may be regularized by spreading it out over a small size $r_0$, 
with $1/r_0$ acting as an ultraviolet cutoff.  In addition, there is
an induced long-range, attractive interaction due to the ``fish'' diagram 
involving exchange of two virtual phions.  This corresponds to a $-a^2/r^3$ 
interaction, if we neglect the mass of the exchanged phions.  (Including the 
phion mass basically cuts off this potential at distances greater than 
$\sim 1/m$.)  

      [For our purposes this form of the phions' interparticle potential is 
effectively exact, provided that $a$ incorporates short-range interactions to 
all orders.  The point is that, just as in the non-relativistic 
theory of BE condensation,\cite{huang} only low-energy ($ka \!\ll\! 1$) 
scattering is involved, and this can be characterized by a single parameter, 
the $s$-wave scattering length $a$.]
  
     Consider a large box, volume $\vol$, with periodic boundary conditions  
that contains $\num$ phions.  Provided the system is dilute 
($n a^3 \!\ll\! 1$), 
the ground state corresponds to almost all the phions being Bose-condensed 
in the zero-momentum mode; hence the energy is
\be
\label{energy}
       E= \num m + \half \num^2 \bar{u},
\ee
where the first term counts the rest energies of the $\num$ phions and the 
second is the number of pairs times the average energy of a pair, $\bar{u}$.  
The diluteness assumption means that three-body interactions, etc., can be 
neglected.  Since almost all the phions are in the zero-momentum mode, whose 
wavefunction is uniform across the box, the average energy of a pair is just
\be
\label{ubar}
      \bar{u} = \frac{1}{\vol} \int \! d^3r \, V(r).
\ee
Substituting this into (\ref{energy}) and dividing by volume gives the energy 
density as 
\be
{\cal E} = n m + n^2 \!\int \! d^3r \, V(r).  
\ee
(I've dropped the $\half$.)  The interparticle potential $V(r)$ contains the 
$(a/m)\d3r$ term and the $-a^2/r^3$ term.  Integration of the latter yields 
$\int \! dr/r$, which is cut off at short distances by the core size $r_0$, 
but will also need to be cut off at long distances by some $r_{\rm max}$:
\be
\label{enden1}
{\cal E} = n m + \frac{n^2 a}{m} - n^2 a^2 \ln \left( \frac{r_{\rm max}}{r_0} 
\right) .
\ee

     The crucial question now is; what determines $r_{\rm max}$?  As mentioned 
earlier, the phion mass will ultimately cut off the $-a^2/r^3$ interaction 
at distances $\geq 1/m$.  However, when $m$ is very small a more important 
consideration is the `screening' by the background density of phions.  
Thus, the $\ln(r_{\rm max})$ will naturally turn into a logarithm of $n$.  

    To see this, consider two, well-separated phions that are trying to 
interact by exchanging a pair of virtual phions.  The two virtual phions
have to travel through the condensate and thus will experience collisions with 
background phions.  These collisions with zero-momentum particles, 
proportional to $n$ times $a$, behave as mass insertions in the propagator.  
They convert each phion propagator into the propagator for a quasiparticle 
(a Higgs boson) whose mass squared is thus
\be 
M^2 = m^2 + 8 \pi n a .
\ee
It turns out that the second term completely dominates, so the Higgs mass 
is much, much greater than the phion mass.  The scale for $r_{\rm max}$ is 
then set by $1/M \sim 1/\sqrt{n a}$ (a much shorter length scale than $1/m$).  
     
      The energy density, from (\ref{enden1}), thus has the form 
\be
{\cal E} = \mbox{{\rm sum of}} \,\, n, \,\, n^2, \,\, n^2 \ln n  \,\,\, 
\mbox{{\rm terms.}}
\ee
The three terms represent (i) one-particle rest energies, (ii) the energy cost 
of repulsive, short-range, two-particle interactions, and (iii) the energy 
{\it gain} due to attractive, long-range, two-particle interactions (with the 
$\ln n$ arising because the incipient infrared divergence is cut off by the 
background density effect).   For sufficiently small phion mass (iii) wins, 
so that an empty box ($n=0$) is energetically disfavoured, compared to a 
condensate of some optimal density $n_v$, where $n=n_v$ is the non-trivial 
minimum of ${\cal E}(n)$.  

      This physics can be translated~\cite{mech} back into familiar 
field-theory terms by recognizing that $n$ translates to $\half m \phi^2$, 
where $\phi$ is the background field value $\langle \Phi \rangle$.  The 
energy density as a function of $n$ then translates into the field-theoretic 
{\it effective potential} as a function of $\phi$. 

\section{Hierarchy of Length Scales}

     It is straightforward to analyze where the phase transition occurs in 
terms of the parameters.~\cite{mech}  In units where the Higgs mass $M_h \sim 
\sqrt{n_v a}$ is finite, we need $m$, $1/n_v$, and $a$ to tend to zero like 
$\epsilon^{1/2}$, where $\epsilon = 1/\ln({\rm cutoff}/M_h)$ and the cutoff is 
$1/r_0$.  The fact that the scattering length $a$ must vanish reflects the 
`triviality' of $\lambda \Phi^4$ theory.  `Triviality' is usually viewed as 
an embarrassment, which is odd because it naturally does something very 
desirable; it generates a hierarchy.  In fact, 

\begin{center}

{\bf `TRIVIALITY'} means {\bf HIERARCHY}. 

\end{center}

\hspace*{-\parindent}
`Triviality' means that there are two, vastly different, physical length 
scales; the Compton wavelength $M_h^{-1}$ (finite) and the scattering length 
$a$ (infinitesimal).  By keeping the cutoff finite, but as large as we like, 
we can have a hierarchy of physical length scales.  In fact, the hierarchy 
is quite  rich, as illustrated below:

%
%   line figure constructed in LaTeX

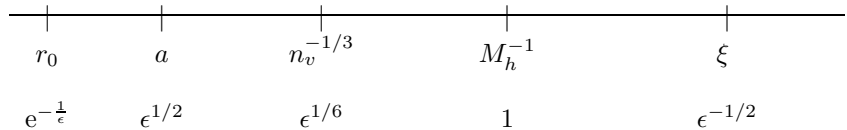
\vspace*{6mm}

\begin{tabular}{ccccc}
\hline
\vspace*{-6mm}
\hspace*{2.0mm} & \hspace*{1.6cm} & \hspace*{1.8cm} & 
\hspace*{2.3cm} & \hspace*{2.7cm} \\
$|$   & $|$   &   $|$          & $|$         &  $|$    \\
$r_0$   & $a$   & $n_v^{-1/3}$            &  $M_h^{-1}$    &   $\xi$  
\vspace*{4mm} \\
${\rm e}^{-\frac{1}{\epsilon}}$ & $\epsilon^{1/2}$  & $\epsilon^{1/6}$ & $1$ & 
$\epsilon^{-1/2}$  
\end{tabular}

\vspace*{4mm}

     The fact that the average phion spacing, $n_v^{-1/3}$, is much greater 
than $a$ corresponds to diluteness.  Both these scales are small compared 
to the physical length scale set by $M_h^{-1}$.  There is also a very 
long length scale, denoted by $\xi$, which corresponds to the phion {\it mfp}, 
$1/(na^2)$, which happens to be the same order as $m^{-1}$.    

     A natural speculation~\cite{mech} is to identify $a$ with the Planck 
length.  In this case $\epsilon$, instead of going to zero, is the tiny 
number $10^{-34}$.  In that case the {\it mfp} is at the 
millimeter/centimeter scale.  Sound waves with wavelengths much longer than 
this scale would correspond to the quasiparticle spectrum having a 
`phonon branch,' $E \sim v_s p$, at ultra-low momenta, $p \ll m$.  This 
`phonon branch' would have to join on to the normal spectrum 
$E \sim \sqrt{p^2+M_h^2}$ at $p \sim m$, implying that the speed of sound 
$v_s$ is of order $M_h/m$, which is much greater than the speed of light 
(by a factor of $10^{17}$ in this scenario).

     Two points should be made about this vastly superluminal sound velocity:
(i) Sound waves exist only for long wavelengths, so there is no way to 
produce a sharp wavefront; thus, superluminal {\it signalling} is still 
impossible.  (ii) The velocity $v_s$ is relative to the condensate rest 
frame, but in a moving frame the apparent sound velocity could be very 
different.  Indeed, in principle, one could use this effect to determine the 
aether rest frame.\footnote{
In the infinite-cutoff limit, where exact Lorentz invariance is recovered, 
the condensate rest frame still exists, but it cannot be determined 
experimentally. 
} 

     There is no space here to discuss some independent reasons to suspect 
that the Higgs spectrum has a `phonon branch' (which would shrink into a 
discontinuity in the propagator at $p^{\mu}p_{\mu}=0$ in the infinite-cutoff 
limit).  I refer the reader to the following 
references:~\cite{consoli,anishetty,cea}  In particular, there are 
longstanding arguments that, as $p \to 0$, the radial Higgs propagator should 
behave as $\mid\!p\!\mid^{d-4}$ in $d$ dimensions.~\cite{anishetty}  There 
is also direct evidence for peculiar infrared behaviour from lattice 
simulations.~\cite{cea} 

\section*{Acknowledgments}
I thank Maurizio Consoli for many discussions about these ideas.   
This work was supported in part by the Department of Energy under Grant No. 
DE-FG05-97ER41031.

\end{document}